\documentclass[pra,twocolumn,showpacs,amsmath,amssymb]{revtex4-1}

\usepackage[pdftex]{graphicx}
\usepackage{dcolumn}
\usepackage{bm}
\usepackage{xcolor}
\usepackage{times}
\usepackage{verbatim}

\newcommand{\bvec}[1]{\mbox{\boldmath$\mathrm{#1}$}}

\begin{document}
\title{Attosecond tracking of light absorption and refraction  in fullerenes}

\author{A.S. Moskalenko}
\email{andrey.moskalenko@physik.uni-halle.de}

\altaffiliation[Also at ]{A.F. Ioffe Physico-Technical Institute,
194021 St. Petersburg, Russia}

\author{Y. Pavlyukh}
\author{J. Berakdar}
\affiliation{Institut f\"ur Physik, Martin-Luther-Universit\"at
 Halle-Wittenberg,
 Heinrich-Damerow-St. 4, 06120 Halle,
 Germany}

\date{\today}
\begin{abstract}
 The collective response of matter
is  ubiquitous and  widely exploited, e.g. in plasmonic, optical
and electronic devices. Here we trace on an attosecond time scale
the birth of collective excitations in a finite system and find distinct new features
in this regime. Combining quantum chemical computation with
{quantum} kinetic methods we calculate the time-dependent light
absorption and refraction in fullerene that serve as indicators for the
emergence of collective modes. We  explain  the numerically calculated  novel transient
features by an analytical model and point out the relevance for ultra-fast photonic and
electronic applications. A scheme is proposed to measure the
predicted effects via the emergent attosecond metrology.

%
%
\end{abstract}


\maketitle
%
%

\section{Introduction}

The last decade has witnessed the emergence of the attosecond
science opening a window  on sub-femtosecond processes  that take
place in atoms and
molecules~\cite{Haessler2010,Shafir2009,Eckle2008,Goulielmakis2008}
(see Refs.~\cite{Corkum2007,Krausz2009,Kling2008,Popmintchev2010}
for reviews). Increased interest is currently focused on many-body
effects and condensed matter systems
\cite{Rohwer2011,Schultze2010,Cavalieri2007} where  many degrees
of freedom may interfere.
 The  hallmark of extended systems is the
collective, dielectric linear response \cite{Mahan_book} that
determines for example the light propagation
\cite{Maier_book,Brongersma2010} and the energy  and momentum loss
of traversing charged particles \cite{Egerton_book,Egerton2009}.
On a fundamental level, the dielectric response describes how the
particles cooperatively act as to screen the interparticle Coulomb
interaction. Obviously, this collective motion builds up on a time
scale on which the effective interparticle  interaction changes
qualitatively (even in sign) \cite{Haug_Jauho}.
 This was impressively demonstrated by terahertz spectroscopy for  a semiconductor-based electron-hole
plasma   \cite{Huber2001} evidencing that the electron-electron interaction develops from
its unscreened to a fully screened form  on a time scale of the
 order of the inverse plasma frequency (several $10^{-14}$ seconds).
For a finite system the situation is qualitatively different: Here
quantum size and topology effects bring about new features. E.g.,
as detailed below  a spherical shell system exhibits two coupled
plasmon modes that, while interfering transiently, evolve to  two
separate features in the long-time limit.
 In addition, these modes lie energetically  in the ultraviolet (UV) or extreme ultraviolet (XUV) energy range.
Hence, tracing their evolution requires sub-femtosecond
resolution, which is within reach experimentally
\cite{Haessler2010,Shafir2009,Eckle2008,Goulielmakis2008,Corkum2007,Krausz2009,Kling2008,Popmintchev2010}.

\begin{figure}
  \includegraphics[width=7.5cm]{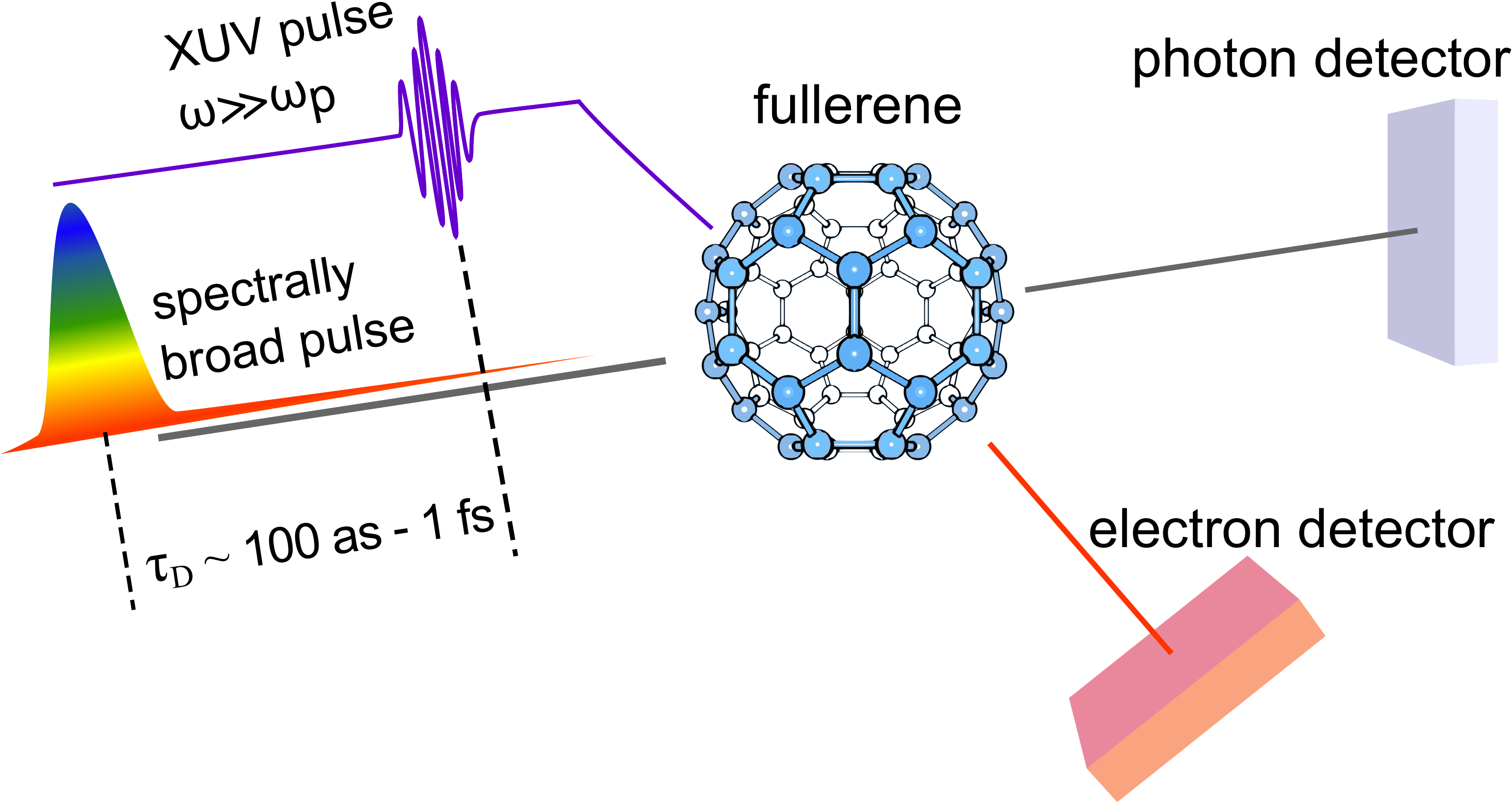}
  \caption{(Color online)
 A schematic of the considered setup. An attosecond XUV pulse ionizes a fullerene.
  After a time delay $\tau_{_{\rm D}}$ the time-dependent absorption of the sample is studied by
analyzing the absorption of broadband, linear-polarized
pulse.\label{Fig:setup}}
\end{figure}

To be definite we concentrate here on the   attosecond ({\it as})
evolution of the dielectric response in finite systems as
monitored by the light absorption and refraction in fullerenes.
The time-non-resolved (we call it hereafter steady-state)
absorption of the fullerene
\cite{Kroto1985,Orlandi2002,Dresselhaus_book}, in our case
C$_{60}$,  can be accessed by analyzing the absorbed components
from an impinging  moderate intensity, broadband linear-polarized
pulse. To monitor the attosecond buildup in absorption,  we
suggest that at a time $t=0$ an XUV attosecond pulse ionizes
C$_{60}$ and the change in the absorption at a time delay
$\tau_{_{\rm D}}$ thereafter is recorded, as sketched in
Fig.~\ref{Fig:setup}.  For negative $\tau_{_{\rm D}}$ we obtain
the absorption of C$_{60}$ and for large (few femtoseconds)
$\tau_{_{\rm D}}$ we expect the steady-state absorption of
C$_{60}^+$. Equilibration in the limit {$t\rightarrow \infty $}
proceeds via electronic, plasmonic and finally ionic channels,
each having a distinct time scale; the ions are frozen at the
\emph{as} timescale. We focus on the plasma frequency
($\omega_{\rm p}$) regime where the absorption and refraction show
strong modulations
 \cite{Bonin_book}.
 The change of $\omega_{\rm p}$ for $t>0$
  is evident from a qualitative
  consideration of  the square root dependence on the density:
  The valence band  of
C$_{60}$  accommodates 240 electrons (originating from  the $2s$ and $2p$ states of each carbon
atom) that form 180 $\sigma$-type and 60 $\pi$-type molecular orbitals
\cite{Orlandi2002}. The highest occupied molecular orbital (HOMO) is five-fold degenerate.
Its complete depletion by the XUV pulse reduces the number of particles  by
4\%, or red-shifts the collective response by 2\%. As theoretically quantified below,
 even a single-photoionization event has a significant
impact on the plasmon dynamics.

\section{Theoretical formulation}

C$_{60}$ {has} a size below 1 nm~\cite{Dresselhaus_book}, its
electronic and optical properties for $t<0$ and {$t\rightarrow
\infty$}  are well-documented \cite{Orlandi2002,Dresselhaus_book}.
Also C$_{60}$ is readily available, very stable, and has an
ionization potential of $\approx$7.5 eV and an electron affinity
of 2 eV. It also exists in a well-studied crystalline form
(fullerite) \cite{Kratschmer1990}. The collective response is
marked by  a giant plasmon resonance at $\approx 22$~eV
\cite{Bertsch1991} that was investigated experimentally in the
solid
 \cite{Hansen1991,Sohmen1992} and the
gas phase \cite{Hertel1992}. Considerable efforts were
devoted  to  clarify quantitatively the experiments~\cite{Bauernschmitt1998,Berkowitz1999}.
It was not until
recently however that the existence of a second plasmon peak at higher
energy ($\approx 39$~eV) but with a lower oscillator strength was
experimentally confirmed~\cite{Reinkoester2004,Scully2005}. Theoretically, the
presence of two collective modes follows from {a
classical dielectric shell
model~\cite{Mukhopadhyay1975,Lambin1992,OEstling1993,Oestling1996,Vasvari1996}}.
The \emph{quantitative} description, as we are seeking here,
 is a challenging task,  however.
Quantum mechanical approaches utilize  either the tight-binding
(TB) model for the valence electrons~\cite{Bertsch1991} or the
jellium shell
model~\cite{Puska1993,Yabana1993,Ivanov2003,Scully2005,Madjet2008}.
The linear response  to an external electric field is calculated
either using the random phase approximation (RPA) for the
polarization propagator ($\chi$) (in the case of the TB
\cite{Bertsch1991} and some of the density-functional-based (DFT)
models \cite{Yabana1993,Ivanov2003}), or {by implementing the
time-dependent DFT}
(TDDFT)~\cite{Puska1993,Scully2005,Madjet2008}. Until now TDDFT
seems to deliver the best agreement with experiments, however, the
theory  describes correctly either the position of the high energy
plasmon resonance or, with the help of an adjustable parameter, the
position of the lower energy plasmon resonance
\cite{Scully2005,Madjet2008}.
Here we develop an approach that rests on three steps: 1. We utilize quantum chemical
\emph{ab-initio} techniques to capture accurately the stationary,
single-particle electronic structure. 2. These states are
expressed in a basis appropriate for many-body calculations which
we perform within the linear response theory (i.e., the random
phase approximation) to obtain the steady-state collective
response. 3. With both quantities at hand we perform quantum
kinetic calculations within the density matrix formalism to obtain
the time and the frequency dependent polarizability
$\alpha(E,t=\tau_{_{\rm D}})$. The imaginary and the real part of
$\alpha(E,t)$ deliver then respectively the time-dependent
absorption and refraction {properties} of the sample
\cite{Bonin_book}.

We showed recently \cite{Pavlyukh2009,Pavlyukh2010,Pavlyukh2011}
that the wave function $\Psi_\epsilon(\vec{r})$ of the valence
band electrons  with an energy {$\epsilon=\varepsilon_{nl}$}, as
obtained
 from first principle calculations,
is expressible to a good approximation as a product of a radial
part $R_n(r)$ (having $n-1$ nodes) and an angular  part described
by spherical harmonics $Y_{lm}(\Omega)$ with $l,m$ being the
orbital and magnetic quantum numbers  ($\vec r \equiv(r,\Omega)$
identifies  the electron position, cf. Ref.~\cite{Suppl_Mat})
\cite{Yabana1993,Pavlyukh2009}. This procedure is shown
\cite{Pavlyukh2009} to be valid for C$_{48}$N$_{12}$, B$_{80}$,
C$_{60}$, C$_{240}$, C$_{540}$, C$_{20}$H$_{20}$, and Au$_{72}$.
Thus, the theory presented below is readily applicable to these
systems; for the sake of clarity the discussion  is restricted to
C$_{60}$, however. For C$_{60}$ the two occupied radial subbands
$\varepsilon_{n=1\, l}$ and $\varepsilon_{n=2\, l}$
 are separated by approximately $17.5~$eV.
 The HOMO-LUMO gap  as
determined by our \textit{ab-initio} calculations  ($E_{\rm g}\approx 5.5$~eV)
  as well as some structural information
 are  encapsulated {in the energy spectrum
 $\varepsilon_{nl}$} (cf.~\cite{Suppl_Mat}).
These data allow us
 to perform the mapping of the unperturbed system to the single-particle Hamiltonian
 $\hat{H}_0$ with the states $\phi_\alpha\equiv\phi_{nlm}=R_n(r)Y_{lm}(\Omega)$ and the
 spectrum $\varepsilon_{nl}$. These
single particle states are then taken as the basis to express  the
density operator {$\hat{\rho}$}.  Our theoretical description of
the charge dynamics is based on the resulting density matrix $
\bvec \rho$.


\begin{figure*}[t]
  \includegraphics[width=15cm]{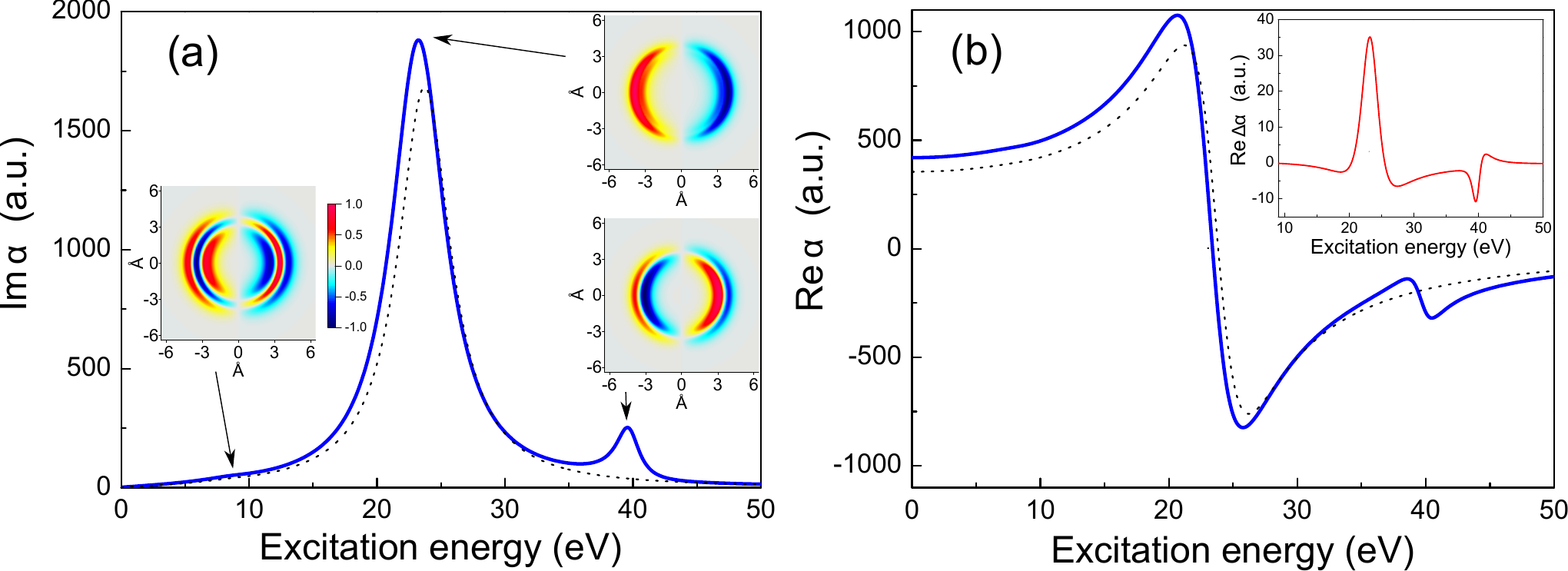}\\
  \caption{(Color online)
  Energy dependence of the (a) real and (b) imaginary parts of the dipolar
  polarizability $\alpha(E)$ of C$_{60}$ in the long-time limit. Insets to (a) show the
  calculated spatial densities of the plasmonic modes, corresponding to the peaks in the
  spectrum, in any plane crossing the center of the molecule and being parallel to the
  polarization of the external electric field. Dotted lines result
  from the approximation including only intraband collective
  excitations. Inset to (b) shows the difference in the
  spectra between C$_{60}$ and C$_{60}^+$ for the case of refraction. \label{Fig:alpha_C60_long_time}}
\end{figure*}


We are interested in two types of external perturbations
$\hat{V}^{\rm ext}$ that trigger the evolution: (i) a broad band,
low-intensity light pulse
probing the plasmonic response and (ii) a sudden change in the
population upon a single ionization of C$_{60}$ by the attosecond
XUV pulse. Formally, the latter change is described by a
time-dependent occupation function $f_{nl}(t)$. To simulate the
plasmonic response, we employ the Heisenberg equation of motion
for the density matrix {$\hat{\rho}$}, the mean-field
approximation and the linear response to the external driving. The
relaxation within a time $\tau$ due to collisions we treat within
the particle-conserving relaxation time approximation
\cite{Mermin1970}. This approach has been successfully tested for
a variety of systems
\cite{Morawetz2005,Huber2001,Borisov2004,Garcia2010,Tanuma2011}.
It should be emphasized, however, that we are interested in the
dynamics at times much shorter than $\tau$. The equation of motion
for  $\hat{\rho}$  reads as
\begin{equation}\label{Eq:rho_HEM}
     \frac{\partial \hat{\rho}}{\partial t}
    +\frac{i}{\hbar}
   \left[\hat{H_0}+\hat{V}^{\rm ind}
    +\hat{V}^{\rm ext},\hat{\rho}\right]
    =\frac{\hat{\rho}^{\rm l.e.}-\hat{\rho}}{\tau}.
\end{equation}
The induced potential $\hat{V}^{\rm ind}$ has to be  determined
self-consistently from the induced charge density as derived from
the change in the density operator $\hat{\rho}$ [to find
$\hat{\rho}$ we need $\hat{V}^{\rm ind}$ in
Eq.~\eqref{Eq:rho_HEM}]. $\hat{\rho}^{\rm l.e.}$ is the local
equilibrium density operator. The {corresponding} locally relaxed
density {$\rho^{\rm l.e.}(\vec{r},t)$} is the distribution that,
at any given instant, would be in equilibrium in the presence of
$\hat{V}^{\rm ext}$ and $\hat{V}^{\rm ind}$ while satisfying the
charge conservation
 ~\cite{Mermin1970}.  Technically,
we express $\hat \rho$ in the basis
{\{}$\phi_\alpha(\vec r)${\}} and expand to first order
around the equilibrium, i.e.
$\rho_{\alpha\beta}(t)=\langle\alpha|\hat{\rho}(t)|\beta\rangle=f_\alpha(t)\delta_{\alpha\beta}+
\delta\rho_{\alpha\beta}(t); \quad
\rho^{\rm l.e.}_{\alpha\beta}(t)= \langle\alpha|\hat{\rho}^{\rm
l.e.}(t)|\beta\rangle=f_\alpha(t)\delta_{\alpha\beta}+
{\pi^0_{\alpha\beta}(t)}\delta\mu_{\alpha\beta}(t),$
where
${\pi^0_{\alpha\beta}(t)}=\frac{f_\alpha(t)-f_\beta(t)}{\varepsilon_\alpha-\varepsilon_\beta}$
and $\delta\bvec\mu(t)$ is the matrix corresponding to the local
chemical potential (full details are given in Appendix
\ref{App:derivation}).

Application of an external linearly-polarized electric field with
an amplitude ${\cal E}_0$, polarization direction $\vec{\rm e}$,
and frequency $\omega$, that corresponds to $\hat{V}^{\rm
ext}(t)=-e{\vec{r}\cdot \vec{\rm e}} {\cal E}_0 e^{-i\omega t}$
($e$ is the electron charge), leads to the change of the density
matrix $\delta\bvec \rho(t)$ which in linear response is governed
by the two-times response function $\bvec \chi(t,t')$, i.e.
$    {\delta \bvec \rho(t)}=\int_{-\infty}^{t}\!\! {\rm d}t'\;\bvec \chi(t,t')
    {\bvec V^\mathrm{ext}(t')}\:. $
The induced dipole moment derives from  $\vec{P}(t)={\rm
Tr}\left[e\vec{r}\delta\hat{\rho}(t)\right]$. For spherical
molecules, reducing all quantities to their radial components, we
find (see Appendix \ref{App:dipolar_polarizability})
\[ P(t)=-\frac{4\pi}{3}e^2{\cal E}_0  \int_{-\infty}^{t}\!\! {\rm
d}t'\; \bvec r^T\bvec\chi(t,t')\bvec r e^{-i\omega t'}{\:,}
\]
%
where the elements of $\bvec r$ are $r_{nn'}=r_{n'n}=\langle n|r|
n'\rangle$. Introducing the dimensionless, Fourier-transformed
response function ($\tilde{\bvec r}={\bvec r}/r_0$, $r_0$ is the
fullerene average radius and  $\epsilon_0$ is the vacuum
permittivity)
\begin{equation}\label{Eq:z_n1n2_def_paper}
    \bvec z(\omega,t)=\frac{e^2}{3\epsilon_0 r_0}
    \int_{0}^{\infty}\!\! {\rm d}\tau \; e^{i\omega \tau}\bvec\chi(t,t-\tau)\tilde{\bvec
    r}\;,
\end{equation}
we find $P(\omega;t)=-4\pi r_0^3\epsilon_0{\cal E}_0{\rm
e}^{-i\omega t}\tilde{\bvec r}^T \bvec z(\omega,t)\; $.
The time and frequency-dependent dipolar
polarizability is then calculated numerically as
\begin{equation}\label{Eq:alpha_via_z_vec}
    \alpha(\omega,t)=- r_0^3\tilde{\bvec r}^T \bvec z(\omega,t)\;.
\end{equation}
(in SI units, $\alpha_{\rm SI}=4\pi\epsilon_0 \alpha$). The
explicit equation for the evolution of $\bvec z(\omega,t)$, and
therefore of $\alpha(\omega,t)$, following from
Eq.~\eqref{Eq:rho_HEM} we included  in Appendix
\ref{App:derivation}.

For an insight into the numerical results we construct an
analytical model based on the following: For C$_{60}$ (and
generally for {,,spherical'' molecules}) we find that the
intraband components {$z_{11}(\omega,t)$} and {$z_{22}(\omega,t)$}
of the response function are hardly influenced by the interband
components {$z_{12}(\omega,t)$} and {$z_{21}(\omega,t)$} (details
are in Appendix \ref{App:approx_stat_sol}).
In this approximation, considering only the intraband terms we
infer a system of two coupled equations for {$z_{11}(\omega,t)$
and $z_{22}(\omega,t)$}
\begin{equation}\label{Eq:z_intraband_ODE}
     \begin{split}
    \Big[\partial^2_t+&\left(\frac{1}{\tau}-\!2i\omega\right)\partial_t
      +\omega_{{\rm F},n}^2\!-\!\omega^2\!-\!i\frac{\omega}{\tau}\Big]z_{nn}(\omega,t)\\
   +&\omega_{{\rm p},n}^2\sum_{n'=1,2}\tilde{g}_{nn,n'n'}z_{n'n'}(\omega,t)
   =-\omega_{{\rm p},n}^2\;,\\
 \end{split}
\end{equation}
where  the stationary matrix $ \tilde{\bvec g}$ is explicitly
defined in Appendix \ref{App:Coulomb} and  $\omega_{{\rm
p},n}^2=1/\hbar^2\times e^2\sqrt{2N_n}/(6\pi\epsilon_0 r_0)\times
\hbar\omega_{{\rm F},n}$. Here $N_n$ is the number of electrons in
the $n$-th radial subband. For C$_{60}$ in the equilibrium we have
$N_1=180$ and $N_2=60$. $\hbar\omega_{{\rm F},n}$ are the energy
distances between the highest occupied state in the $n$-th radial
subband and the unoccupied state in the same subband having the
value of $l$ that is greater by one. On the same approximation
level Eq.~\eqref{Eq:alpha_via_z_vec} simplifies to
$\alpha(\omega,t)=
-r_0^3\left[z_{11}(\omega,t)+z_{22}(\omega,t)\right]$ meaning that
the interband components can be then neglected when calculating
$\alpha(\omega,t)$. In such a consideration the dynamics of the
electron density in two different radial channels are still
coupled via the induced fields like in concentric
nanoshells~\cite{Radloff2004}.

\section{Steady-state collective response}

Absorption and refraction spectra of the C$_{60}$  and C$_{60}^+$
in the long-time limit (i.e. $\tau_{_{\rm D}}<0$) are shown in
Fig.~\ref{Fig:alpha_C60_long_time}. The responses of C$_{60}^+$
and C$_{60}$ are quite similar [cf. inset in
Fig.~\ref{Fig:alpha_C60_long_time}(b)] which is in agreement with
the existing theory and experiment \cite{Scully2005,Madjet2008}.
 Our  C$_{60}$ calculations exhibit two peaks at 23.2~eV and
 at 39.6~eV  in the absorption spectrum which correspond to plasmon excitations
and  agree very well with  experiment \cite{Scully2005}. The width of the measured main plasmon resonance
 is reported to be  $\approx 5$~eV  which sets
the scale for the relaxation time $\tau$. {With
$\sigma(E)=\frac{4\pi E}{\hbar c}{\rm Im}\alpha(E)$
\cite{Bonin_book}, where $c$ is the speed of light, we} find the
corresponding {absorption} cross-sections $\sigma(E)$ to be
$4.1\times 10^3$~Mb for the main peak and $9.4\times 10^2$~Mb for
the higher energy peak. {This} yields relative heights of the
peaks that {are} in qualitative agreement with the experiments and
with the time-dependent DFT results \cite{Puska1993,Madjet2008}.
There is also a weak low energy  contribution at
$\approx$ 9~eV. Such structures in the absorption
have been  attributed to
single-electron transitions \cite{Bauernschmitt1998,Orlandi2002}.
Our calculated resonance has a collective character and is
energetically in the proximity of the single-electron excitations
making it difficult  for an observation via absorption. As
well-established \cite{Bonin_book}, the peaks in the absorption
(refraction) are symmetric  (antisymmetric) with respect to
reflection at the respective central frequency.

\begin{figure*}[t]
\centering
  \includegraphics[width=\textwidth]{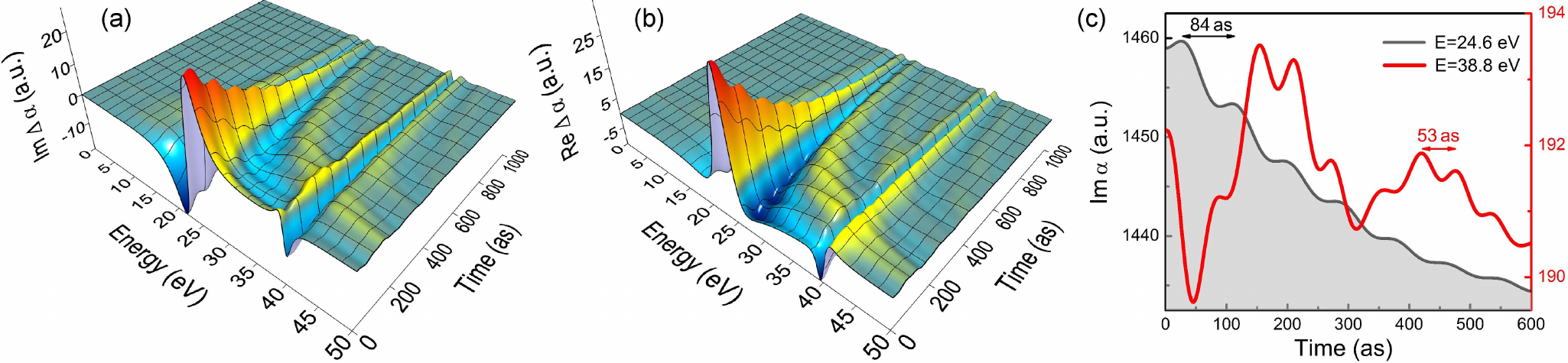}
  \caption{(Color online) (a) Difference between the imaginary part of the dipolar polarizability ${\rm
  Im}\alpha(E,t)$ after removing one electron from the highest occupied state of C$_{60}$
  in the second (upper) branch at $t=0$ and its long-time limit for C$_{60}^+$, i.e. ${\rm
  Im}\alpha_{{\rm C}_{60}^+}(E)$.  (b) The same dependence for the real part change of
  $\alpha(E,t)$. (c) Time dependence of the imaginary part of the dipolar polarizability $\mbox{Im}\hspace{2pt}\alpha(E,t)$ of the excited
  fullerene molecule at fixed energy values, $E=24.6~$eV and $E=38.8~$eV, taken close to the main and higher energy
  resonance positions of the stationary response, respectively. Time intervals $T=\pi\hbar/E$ indicated by the horizontal double-arrow
  lines are close to the observed oscillation periods for each of the
  dependencies: $T=84$~{\it as}  and $T=53$~{\it as} , respectively.}
  \label{Fig:dalpha_C60plus}
\end{figure*}


The approximation including only intraband collective excitations,
i.e. considering the stationary solution of
Eq.~\eqref{Eq:z_intraband_ODE}, gives a very good description of
the main peak in the spectrum (see
Fig.~\ref{Fig:alpha_C60_long_time}), but misses the higher energy
plasmon. The last is caused by interband collective excitations
(i.e. governed by $z_{12}$ and $z_{21}$) that are still influenced
strongly by the intraband response.
The interband excitations  shift   slightly the main plasmon
resonance to lower energies and enhance  its strength.
The $f$-sum rule \cite{Apell1993,Ivanov2003, Pal2011} for the  absorption
 is fulfilled only approximately, particularly because higher energy
 states in the continuum and close to it are not included in our model. However,
these states   contribute  mostly to the background of the
spectrum.
For a further  insight we show the calculated
 spatial distributions of the induced charge densities
at the maximal absorption [insets in
Fig.~\ref{Fig:alpha_C60_long_time}(a)].
The oscillation amplitude
depends linearly on the external field strength. The
interband character of the the higher energy plasmon excitation is evident.
Its radial structure is determined by the product $R_1(r)R_2(r)$ and has therefore  one node
located in the middle of the fullerene  cage. The
angular dependence of the modes in and out of the depicted plane
 is governed  by their dipolar character. The
radial structure of the density oscillation at  the
main plasmon peak shows  a constructive superposition of the radial
electronic radial density of the first $|R_1(r)|^2$ and
the second $|R_2(r)|^2$ radial channels that oscillate in phase.
In contrast, we observe an out-of-phase oscillation  of
the densities in different channels at 9~eV. Both
cases are collective intraband electronic excitations.

\section{Transient dynamics}

Having   assessed successfully the steady-state response we move
on to the attosecond transient dynamics.  We focus on the setup
shown in Fig.~\ref{Fig:setup}. At $t=0$ the XUV pulse with energy
$\omega \gg \omega_{\rm p}$
 and  a duration   $<100$~{\it as}
 changes  the population swiftly in  C$_{60}$
by ejecting one electron.
To access the transient  absorption or refraction
let  us assume the photoelectron is emitted at $t=0$  from
the second radial band leaving C$_{60}^+$  (we find similar results for
 multiple XUV ionization). For all
time moments before the ionization the time-dependent response
$\alpha(E,t<0)$ is equal to its steady-state value $\alpha_{{\rm
C}_{60}}(E)$ for C$_{60}$. At $t>0$ we solve for the dynamics of
$\alpha(E,t)$, which for long times approaches the steady-state
value $\alpha_{{\rm C}_{60}^+}(E)$  for C$_{60}^+$.
 The results  in
Figs.~\ref{Fig:dalpha_C60plus}(a) and \ref{Fig:dalpha_C60plus}(b)
illustrate the change in the imaginary (absorption) and real
(refraction) parts of the polarizability  by evaluating
{$\Delta\alpha(E,t)= \alpha_{{\rm C}_{60}}(E,t)-\alpha_{{\rm
C}_{60}^+}(E)$}. The transient dynamics shows a rich structure
evolving over approximately 2~fs. {For $1$ fs the remainder of the
vanishing difference between $\alpha(E,t)$ and $\alpha_{{\rm
C}_{60}^+}(E)$ is still visible in
Figs.~\ref{Fig:dalpha_C60plus}(a) and
\ref{Fig:dalpha_C60plus}(b).} For $t<100$~{\it as}  the response
is basically determined by that of C$_{60}$, i.e. it takes around
100~\textit{as} for C$_{60}^+$ to start responding collectively
and it attains its full, steady state response for $t> 1$~fs. The
reason of this inertia is the finite mass of the carriers. Three
marked general transient features can be distilled from
Fig.~\ref{Fig:dalpha_C60plus}. The dispersive hump starting around
200~{\it as}  is due to the sudden change in the population (and
hence the wide frequency perturbation) brought about by the
shortness of the XUV pulse. Furthermore, in addition to
 the relaxation to the steady-state response we observe
an oscillatory behavior in time with a certain frequency. This behavior
 is most obvious around $\omega_{\rm p}$ (cf. Fig.~\ref{Fig:dalpha_C60plus}).
 The origin of these {oscillations} is comprehensible from a consideration of
 the response of the
fullerene molecule  in the lowest approximation: As we
demonstrated by Eq.~\eqref{Eq:z_intraband_ODE} the response
resembles two coupled damped harmonic oscillators.
 For photon
frequencies $\omega$ close to the main peak position the model can
even be further simplified and the spectra are determined approximately
by the dynamics of a single driven damped harmonic oscillator with
the frequency corresponding to the main peak position. Hence, the feature in the response
we can now understand from the known properties of the driven, damped harmonic motion.
 On a short
time scale this dynamics is  governed  by a combined decay and
oscillations with a frequency around $2\omega$, which is clearly
observed in the full-fledge calculations shown in
{Fig.~\ref{Fig:dalpha_C60plus}(c)} for
$\mbox{Im}\hspace{2pt}\alpha(E,t)$ at $E=24.6~$eV (time period
corresponding to $2\omega$ is indicated by a horizontal
double-arrow line for comparison). The dynamics of
$\mbox{Im}\hspace{2pt}\alpha(E,t)$ for fixed energy $E=38.8~$eV in
the vicinity of the higher energy peak is contributed to by more
frequencies because the higher energy plasmon is strongly
influenced by the main plasmon. However, oscillations
corresponding to $2\omega$ are also seen.

\section{Summary}
We developed a framework for the attosecond
collective response in finite systems with spherical symmetry and
applied
 it to fullerenes. The predicted marked features in the transient absorption
and refraction  should be accessible with current attosecond
metrology. The {discovered} attosecond dynamics in the
optical response points to new opportunities for optoelectronic
devices at the sub-femtosecond time scale.

\section*{Acknowledgements}
It is a pleasure to acknowledge   clarifying discussions on the
experiments with  E. Goulielmakis and R.
Ernstorfer.

\appendix

\section{Derivation of the linear response equations}\label{App:derivation}

The transient  time and  energy-dependent polarizability
$\alpha(t, E)$  is governed, to a first order (linear response),
by the two-time response function $\chi(t,t')$. For the
determination of $\chi(t,t')$ we setup a calculational scheme
based on the solution
 of  the Heisenberg equation
of motion for the density matrix $\hat{\rho}$ under the influence
of the external field $\hat{V}^{\rm ext}$ in the mean-field
approximation that leads to Eq.~\eqref{Eq:rho_HEM}.
In the basis $\{\phi_\alpha\}$ we find the density matrix elements
as $\rho_{nlm,n'l'm'}=\langle nlm|\hat{\rho}|n'l'm'\rangle$. From
the change in the density matrix $\delta\rho_{nlm,n'l'm'}$ we
calculate the change in the charge  density
\begin{equation}\label{Eq:delta_n_full}
    \delta n(\vec{r})=\sum_{lm} \delta n_{lm}(r) Y_{lm}(\theta,\phi)\;,
\end{equation}
where
\begin{equation}\label{Eq:delta_n_lm}
    \delta n_{lm}(r)=\sum_{n'n''} s_{n'n''}(r) \delta n^{n'n''}_{lm}\;,
\end{equation}
\begin{equation}\label{Eq:s_nn}
     s_{n'n''}(r)=R_{n'}(r)R_{n''}(r)\;,
\end{equation}
\begin{equation}\label{Eq:delta_n_nn_lm}
     \delta n^{n'n''}_{lm}=\sum_{l'm',l''m''} y^{l'm'}_{lm,l''m''}
     \delta\rho_{nlm,n'l'm'}\;,
\end{equation}
\begin{equation}\label{Eq:y_lm_lm_lm}
 \begin{split}
  y^{lm}_{l'm',l''m''}&=\int\!\! {\rm d}\Omega\;
  Y_{lm}^*(\Omega)Y_{l'm'}(\Omega)Y_{l'm'}(\Omega)\\
  &=\sqrt{\frac{(2l'+1)(2l''+1)}{4\pi(2l+1)}}\; C^{l0}_{l'0,l''0} C^{lm}_{l'm',l'' m''} \; .
  \end{split}
\end{equation}
 $C^{lm}_{l'm',l'' m''}$ are   Clebsch-Gordan
coefficients. The change in the charge density determines the
induced potential $\hat{V}^{\rm ind}$ leading thus to a
self-consistent procedure.

In the basis $\{\phi_\alpha\}$ the Heisenberg equation of motion
for the density matrix \eqref{Eq:rho_HEM} reads:
\begin{equation}\label{Eq:rho_HEM_basis}
  \begin{split}
     \frac{\partial \rho_{nlm,n'l'm'}}{\partial
     t}=&-\frac{1}{\hbar}\left[i(\varepsilon_{nlm}-\varepsilon_{n'l'm'})+\frac{\hbar}{\tau}\right]\rho_{nlm,n'l'm'}\\
     &-\frac{i}{\hbar}\sum_{n'' l'' m''}
     \left(\rho_{n''l''m'',n'l'm'}V_{nlm,n''l''m''}\right.\\
          &\hspace{1.8cm} \left.-\rho_{nlm,n''l''m''}V_{n''l''m'',n'l'm'}\right)\\
     &+\frac{\rho_{nlm,n'l'm'}^{\rm l.e.}-\rho_{nlm,n'l'm'}}{\tau}
    \;.
  \end{split}
\end{equation}
Here $\varepsilon_{nlm}$ are the single particle energies (cf.
Ref.~\cite{Suppl_Mat}). The matrix elements of the Coulomb
potential are cast as
\begin{equation}\label{Eq:V}
\begin{split}
    V_{nlm,n'l'm'}&=\langle nlm|\hat{V}^{\rm ind}+\hat{V}^{\rm
    ext}|n'l'm'\rangle\\
     &=V^{\rm ind}_{nlm,n'l'm'}+V^{\rm ext}_{nlm,n'l'm'}\;,
\end{split}
\end{equation}
where
\begin{equation}\label{Eq:V_ext}
     V^{\rm ext}_{nlm,n'l'm'}=
     \sum_{l''m''}y^{lm}_{l'm',l''m''}
     V^{{\rm ext},nn'}_{l''m''}\;
\end{equation}
and
\begin{equation}\label{Eq:V_ext,nn}
     V^{{\rm ext},nn'}_{l'' m''}=\int\!\! r^2{\rm d}r
     s_{nn'}(r)V_{l''m''}^{\rm ext}(r)\;.
\end{equation}
Upon
solving the Poisson equation (see Appendix~\ref{App:Coulomb}) we
find for the matrix elements of the potential $V^{\rm
ind}(\vec{r})=e\Phi(\vec{r})$ induced by the change in the charge
density $\delta n(\vec{r})$ the following expression
\begin{equation}\label{Eq:V_ind}
     V^{\rm ind}_{nlm,n'l'm'}=\frac{e^2}{\epsilon_0}\sum_{n_1 n_2}
    \sum_{l''m''}y^{lm}_{l'm',l'' m''}
     g^{l''}_{nn',n_1 n_2}\delta
     n^{n_1,n_2}_{l''m''}\;.
\end{equation}
The coefficients $g^{l''}_{nn',n_1 n_2}$ are defined in
Appendix~\ref{App:Coulomb} by Eq.~\eqref{Eq:g_nn_nn_l_def}.

We are interested in the linear response and expand therefore the
density matrix around the equilibrium distribution $f_{nlm}^0$ as
\begin{equation}\label{Eq:rho_linear_response}
  \rho_{nlm,n'l'm'}=f_{nlm}^0+\delta\rho_{nlm,n'l'm'}\; .
\end{equation}
Thus, $\delta\rho_{nlm,n'l'm'}$ is a small perturbation. The
elements of the local equilibrium density matrix including the
lowest order correction with respect to the equilibrium density
matrix are given then by \cite{Mermin1970}
\begin{equation}\label{Eq:rho_le_elements}
 \begin{split}
  \rho_{nlm,n'l'm'}^{\rm
  l.e.}=&f_{nlm}^0\delta_{nn'}\delta_{ll'}\delta_{mm'}\\
  &-\frac{f_{nlm}^0-f_{n'l'm'}^0}{\varepsilon_{nlm}-\varepsilon_{n'l'm'}}\delta\mu_{nlm,n'l'm'}\;.
  \end{split}
\end{equation}
Here the components of the chemical potential can be expressed as
\begin{equation}\label{Eq:delta_mu}
      \delta\mu_{n'l'm',n''l''m''}=
     \sum_{\tilde{l}\tilde{m}}y^{l'm'}_{l''m'',\tilde{l}\tilde{m}}
     \delta\mu^{n'n''}_{\tilde{l}\tilde{m}}\;.
\end{equation}
We require  density conservation, meaning that
$n^{n'n''}_{lm}=n^{{\rm l.e.},n'n''}_{lm}$ which in turn leads to
\begin{equation}\label{Eq:n_nn_lm}
      \delta n^{n'n''}_{lm}=
     \sum_{\tilde{l}\tilde{m}} M^{n'n''}_{lm,\tilde{l}\tilde{m}}
     \delta\mu^{n'n''}_{\tilde{l}\tilde{m}}\;,
\end{equation}
where
\begin{equation}\label{Eq:M_nn_lm_lm}
      M^{n'n''}_{lm,\tilde{l}\tilde{m}}=-\!\!\!\!\sum_{l'm',l''m''} y^{l'm'}_{lm,l''m''}
      y^{l'm'}_{l''m'',\tilde{l}\tilde{m}}
      \frac{f^0_{n'l'm'}-f^0_{n''l''m''}}{\varepsilon_{n'l'm'}-\varepsilon_{n''l''m''}}\;.
\end{equation}
In general, $\delta\mu^{n'n''}_{\tilde{l}\tilde{m}}$ is determined
from Eq.~\eqref{Eq:n_nn_lm} by  solving a system of linear
equations for each pair of indexes $n',n''$. This step of the
calculation is significantly simplified by neglecting the energy
splitting of the multiplet with the same $l$, i.e. by writing
$\varepsilon_{nlm}=\varepsilon_{nlm'}\equiv \varepsilon_{nl}$ and
therefore $f^0_{nlm}=f^0_{nlm'}\equiv f^0_{nl}$, which is a good
approximation for ``spherical'' molecules such as C$_{60}$, as
confirmed by the ab-initio calculations. Adopting this
approximation we conclude  that (cf. Appendix  \ref{App:chem})
\begin{equation}\label{Eq:mu_nn_simplified}
     \delta\mu^{n'n''}_{lm}= \frac{1}{M^{n'n''}_l}\delta n^{n'n''}_{lm}\;,
\end{equation}
where
\begin{equation}\label{Eq:M_nn_l}
      M^{n'n''}_{l}=-\sum_{l'l''} K^{l}_{l'l''}
      \frac{f^0_{n'l'}-f^0_{n''l''}}{\varepsilon_{n'l'}-\varepsilon_{n''l''}}\;
\end{equation}
and
\begin{equation}\label{Eq:K}
      K^{l}_{l'l''}=\frac{(2l'+1)(2l''+1)}{4\pi(2l+1)}\left[C^{l0}_{l'0,l''0}\right]^2\;.
\end{equation}
Gathering the informations in Eq.~\eqref{Eq:delta_mu} and
Eq.~\eqref{Eq:mu_nn_simplified} we can state
\begin{equation}\label{Eq:delta_mu_nlm_nlm}
      \delta\mu_{n'l'm',n''l''m''}=
     \sum_{\tilde{l}\tilde{m}}y^{l'm'}_{l''m'',\tilde{l}\tilde{m}}
     \frac{1}{M^{n'n''}_{\tilde{l}}}\delta n^{n'n''}_{\tilde{l}\tilde{m}}\;.
\end{equation}

From  Eq.~\eqref{Eq:rho_HEM_basis} we deduce then the following
 determining equation for the
 change in the density matrix
\begin{widetext}

\begin{equation}\label{Eq:rho_HEM_linear_responce}
  \begin{split}
     \frac{\partial \delta\rho_{nlm,n'l'm'}}{\partial
     t}=&-\frac{1}{\hbar}\left[i(\varepsilon_{nl}-\varepsilon_{n'l'})+\frac{\hbar}{\tau}\right]\delta\rho_{nlm,n'l'm'}\\
     &+\frac{i}{\hbar}(f^0_{nl}-f^0_{n'l'})V_{nlm,n'l'm'}-\frac{1}{\tau}
      \frac{f^0_{nl}-f^0_{n'l'}}{\varepsilon_{nl}-\varepsilon_{n'l'}}\delta\mu_{nlm,n'l'm'}
    \;.
  \end{split}
\end{equation}
The solution of this equation is found as
\begin{equation}\label{Eq:rho_HEM_solution}
\begin{split}
  \delta\rho_{nlm,n'l'm'}(t)=\frac{i}{\hbar}\int_{-\infty}^t\!\!{\rm d}t'
     e^{\left[\frac{i}{\hbar}(\varepsilon_{nl}-\varepsilon_{n'l'})+\frac{1}{\tau}\right](t-t')}
      \bigg\{ & \left[f^0_{nl}(t')-f^0_{n'l'}(t')\right]V_{nlm,n'l'm'}(t')\\
      & +i\frac{\hbar}{\tau}
      \frac{f^0_{nl}(t')-f^0_{n'l'}(t')}{\varepsilon_{nl}-\varepsilon_{n'l'}}\delta\mu_{nlm,n'l'm'}(t')\bigg\}\;.
\end{split}
\end{equation}
Inserting  Eqs.~\eqref{Eq:V_ind},\eqref{Eq:V_ext} and
\eqref{Eq:delta_mu_nlm_nlm} into Eq.~\eqref{Eq:rho_HEM_solution}
and summing the left and right hand sides weighted by
$y^{l'm'}_{lm,l''m''}$ as in Eq.~\eqref{Eq:delta_n_nn_lm} we infer
that
\begin{equation}\label{Eq:delta_n_final}
   \begin{split}
      \delta n^{n'n''}_{lm}=&\int_{-\infty}^t\!\! {\rm d}t'\; \Pi^{n'n''}_l(t,t')V^{{\rm ext},n'n''}_{lm}(t')
      +\int_{-\infty}^t\!\! {\rm d}t'\; \Pi^{n'n''}_l(t,t')\frac{e^2}{\epsilon_0}
      \sum_{n_1n_2}g^{l}_{n'n'',n_1n_2}\delta
      n^{n_1n_2}_{lm}(t')\\
      &+\int_{-\infty}^t\!\! {\rm d}t'\; I^{n'n''}_l(t,t')\delta n^{n'n''}_{lm}(t')
      \;,
   \end{split}
\end{equation}
where
\begin{equation}\label{Eq:Pi_tt}
      \Pi^{n'n''}_l(t,t')=\frac{i}{\hbar}\sum_{l'l''} K^l_{l'l''}
      e^{\left[\frac{i}{\hbar}(\varepsilon_{n'l'}-\varepsilon_{n''l''})+\frac{1}{\tau}\right](t'-t)}
      \left[f^0_{n'l'}(t')-f^0_{n''l''}(t')\right]\;,
\end{equation}
\begin{equation}\label{Eq:I_tt}
      I^{n'n''}_l(t,t')=-\frac{1}{\tau M_l^{n'n''}}\sum_{l'l''} K^l_{l'l''}
      e^{\left[\frac{i}{\hbar}(\varepsilon_{n'l'}-\varepsilon_{n''l''})+\frac{1}{\tau}\right](t'-t)}
      \frac{f^0_{n'l'}(t')-f^0_{n''l''}(t')}{\varepsilon_{n'l'}-\varepsilon_{n''l''}}\;.
\end{equation}
This is the central integral equation, which has to be solved to
arrive at the time-dependent density change induced by a
perturbing external electric field.

We introduce  the  response function $\chi^l_{nn',n_1 n_2}(t,t')$
as
\begin{equation}\label{Eq:chi_general_def}
    \delta n^{n' n''}_{lm}(t)=\int_{-\infty}^{t}\!\! {\rm d}t'\;
    \sum_{n_1n_2}\chi^l_{n' n'',n_1n_2}(t,t')V_{lm}^{{\rm ext},n_1 n_2}(t')\;.
\end{equation}
Here  we consider the perturbation by the external electric field
in the dipole approximation. Therefore the calculation can be
limited to the case $l=0$. In order to simplify notations, in the
following we omit the corresponding index by all variables and
coefficients.

Inserting the definition \eqref{Eq:chi_general_def} into
Eq.~\eqref{Eq:delta_n_final}, changing the order of integration in
the second term on the right hand side, comparing again with
Eq.~\eqref{Eq:chi_general_def} and making use of the dipole
approximation, we get the following integral equation for the
response function:
\begin{equation}\label{Eq:chi_general_integral_eq}
\begin{split}
    \sum_{n_1n_2}\chi_{n'
    n'',n_1n_2}(t,t')\tilde{r}_{n_1n_2}=&\Pi^{n'n''}(t,t')\tilde{r}_{n'n''}\\
    &+\int_{t'}^t\!\!{\rm d}t''\;
    \frac{e^2}{\epsilon_0}\Pi^{n'n''}(t,t'')
    \sum_{n_1n_2,n_3n_4}g_{n'n'',n_1n_2}\chi_{n_1n_2,n_3n_4}(t'',t')\tilde{r}_{n_3n_4}\\
    &+\int_{t'}^t\!\!{\rm d}t''\;
    I^{n'n''}(t,t'')\sum_{n_1n_2}\chi_{n'n'',n_1n_2}(t'',t')\tilde{r}_{n_1n_2}\;,
\end{split}
\end{equation}
where the matrix elements $\tilde{r}_{nn'}$ are given by $
\tilde{r}_{n'n''}=\frac{1}{r_0}\int\!\! r^2 {\rm d}r\:
s_{n'n''}(r) r\;$ and are calculated by us using the radial
functions for C$_{60}$ from our \textit{ab-initio} calculations.
Here we have also taken into account that in the dipole
approximation the identity $V^{{\rm ext},n_1
n_2}\tilde{r}_{n'n''}=V^{{\rm ext},n' n''}\tilde{r}_{n_1n_2}$
holds. The corresponding time- and frequency-dependent response
function is determined then by
\begin{equation}\label{Eq:z_n1n2_def}
    z_{n'n''}(\omega,t)=\frac{e^2 g}{\epsilon_0}
    \int_{0}^{\infty}\!\! {\rm d}T\; e^{i\omega
    T}\sum_{n_1,n_2}\chi_{n'n'',n_1n_2}(t,t-T)\tilde{r}_{n_1n_2}\;,
\end{equation}
where $g=1/(3r_0)$ and $\epsilon_0$ is the vacuum permittivity.
Here $r_0\approx 6.745 a_{_{\rm B}}$ is the average radius of the
C$_{60}$ atomic cage, where $a_{_{\rm B}}$ is the Bohr radius.
After inserting Eq.~\eqref{Eq:chi_general_integral_eq} into
Eq.~\eqref{Eq:z_n1n2_def} we arrive at the following system of
integral equations for each frequency value $\omega$:
\begin{equation}\label{Eq:z_n1n2_dynamics}
    \begin{split}
    z_{n'n''}\!(\omega,t)=&\int_{-\infty}^t\!\!{\rm d}t'
    e^{i\omega(t-t')}\tilde{r}_{n'n''} W^{n'n''}(t,t')\\
    &+\int_{-\infty}^t\!\!{\rm d}t' e^{i\omega(t-t')}
    W^{n'n''}(t,t')\sum_{n_1, n_2}\tilde{g}_{n'n'',n_1n_2}
    z_{n_1n_2}(\omega,t')\\
    &+\int_{-\infty}^t\!\!{\rm d}t' e^{i\omega(t-t')}
    I^{n'n''}(t,t')z_{n'n''}\!(\omega,t')\;,
    \hspace{0.4cm} n',n''=1,2\; ;
    \end{split}
\end{equation}
\end{widetext}
where $\tilde{g}_{n'n'',n_1n_2}=g_{n'n'',n_1n_2}/g$ and
\begin{equation}\label{Eq:W^n1n2_tt}
      W^{n'n''}(t,t')=\frac{e^2g}{\epsilon_0}\Pi^{n'n''}(t,t')\;.
\end{equation}
Here $\Pi^{n'n''}$ and $I^{n'n''}(t,t')$ are given by
Eq.~\eqref{Eq:Pi_tt} and Eq.~\eqref{Eq:I_tt}, respectively, with
$l=1$. The system of equations \eqref{Eq:z_n1n2_dynamics} is
solved numerically.

It can be shown (see section \ref{App:dipolar_polarizability})
that the time- and frequency-dependent dipolar polarizability can
be written as
\begin{equation}\label{Eq:alpha_via_zn1_n2}
    \alpha(\omega,t)=- r_0^3\sum_{n'n''} \tilde{r}_{n'n''} z_{n'n''}(\omega,t)\;.
\end{equation}

\section{Calculation of the induced potential and its matrix
elements}\label{App:Coulomb}
Here we determine the electrostatic potential $\Phi(\vec{r})$
induced by the change of the density in the spherical layer
$n(\vec{r})$. The Poisson equation
\begin{equation}\label{Eq:Poisson}
  \Delta\Phi(\vec{r})=-\frac{e}{\epsilon_0}n(\vec{r})
\end{equation}
has then to be solved (SI units are used in this work). Its
solution can be written using the Green's function
$G(\vec{r},\vec{r}')$
\begin{equation}\label{Eq:Phi_Green}
  \Phi(\vec{r})=-\frac{e}{\epsilon_0}\int G(\vec{r},\vec{r}')
  n(\vec{r}'){\rm d}^3\vec{r}',
\end{equation}
where
\begin{equation}\label{Eq:Green}
  G(\vec{r},\vec{r}')=-\frac{1}{4\pi\left|\vec{r}-\vec{r}'\right|}\;.
\end{equation}
Using the spherical geometry of the problem we apply the following
decomposition for the Green's function:
\begin{equation}\label{Eq:one_over_r_decomposition}
  \frac{1}{\left|\vec{r}-\vec{r}'\right|}=
  \sum_{l=0}^\infty\sum_{m=-l}^l\frac{4\pi}{2l+1}\frac{r_<^l}{r_>^{l+1}}Y_{lm}(\theta,\phi)Y^*_{lm}(\theta',\phi')\;,
\end{equation}
where $r_>={\rm max}\{r,r'\}$ and $r_<={\rm min}\{r,r'\}$. Then we
get the following expression for the potential
\begin{equation}\label{Eq:Phi_Green_decomposition}
\begin{split}
  \Phi(\vec{r})=\frac{e}{\epsilon_0}\int {\rm d}^3\vec{r}'
  \sum_{l=0}^\infty\sum_{m=-l}^l &\frac{1}{2l+1}\frac{r_<^l}{r_>^{l+1}}
   Y_{lm}(\theta,\phi)\\
   &\times Y^*_{lm}(\theta',\phi')
  n(\vec{r}')\;.
\end{split}
\end{equation}

To proceed further we decompose $n(\vec{r}')$ in spherical
harmonics
\begin{equation}\label{Eq:n_r_theta_phi}
  n(r,\theta,\phi)=\sum_{lm} n_{lm}(r) Y_{lm}(\theta,\phi),
\end{equation}
where the radial-dependent angular components of the density can
be found as
\begin{equation}\label{Eq:n_lm}
    n_{lm}(r)=\sum_{n'n''} s_{n'n''}(r) n^{n'n''}_{lm}\;,
\end{equation}
with
\begin{equation}\label{Eq:s_nn_app}
     s_{n'n''}(r)=R_{n'}(r)R_{n''}(r)\;,
\end{equation}
and finally expressed via the components of the density matrix
$\rho_{nlm,n'l'm'}$ because
\begin{equation}\label{Eq:n_nn_lm_rho}
     n^{n'n''}_{lm}=\sum_{l'm',l''m''} y^{l'm'}_{lm,l''m''}
     \rho_{nlm,n'l'm'}\; .
\end{equation}
After inserting Eq.~\eqref{Eq:n_r_theta_phi} with
Eq.~\eqref{Eq:n_lm} into Eq.~\eqref{Eq:Phi_Green_decomposition} we
get
\begin{equation}\label{Eq:Phi_final}
  \Phi(\vec{r})=\frac{e}{\epsilon_0}\sum_{nn'}
  \sum_{l=0}^\infty\sum_{m=-l}^l n_{lm}^{nn'} I^{nn'}_l(r) Y_{lm}(\theta,\phi)\;,
\end{equation}
where
\begin{equation}\label{Eq:I_nn}
  I^{nn'}_l(r)=\frac{1}{2l+1}\int\!\! {\rm d}r'\;
  r'^2\frac{r_<^l}{r_>^{l+1}}s_{nn'}(r')\;.
\end{equation}
The determination of the induced electrostatic potential
\eqref{Eq:Phi_final} entails the knowledge of the elements of the
density matrix \eqref{Eq:n_nn_lm_rho}.

We calculated the matrix elements $\langle
n'l'm'|\Phi(\vec{r})|n''l''m''\rangle\equiv\Phi_{n'l'm',n''l''m''}$
of the induced potential as
\begin{equation}\label{Eq:Phi_matrix_elements}
     \Phi_{n'l'm',n''l''m''}=\frac{e}{\epsilon_0}\sum_{n_1 n_2}
     \sum_{\tilde{l}\tilde{m}}y^{l'm'}_{l''m'',\tilde{l}\tilde{m}}
     g^{\tilde{l}}_{n'n'',n_1 n_2}n^{n_1,n_2}_{\tilde{l}\tilde{m}}\;,
\end{equation}
where $y^{l'm'}_{l''m'',\tilde{l}\tilde{m}}$ is given by
Eq.~\eqref{Eq:y_lm_lm_lm}
and
\begin{equation}\label{Eq:g_nn_nn_l_def}
   g^{l}_{n'n',n_1 n_2'}=\int\!\! {\rm d}r\; r^2
   s_{n'n''}(r)I_{l}^{n_1 n_2}(r)\;.
\end{equation}
The values of all matrix elements entering
Eq.~\eqref{Eq:g_nn_nn_l_def} are obtained numerically using the
radial functions for C$_{60}$ from our \textit{ab-initio}
calculations.

As explained above, for  C$_{60}$ we  may adopt the  approximation
of a spherical layer with
 the width of the
 layer $d$ being   smaller than the sphere radius $d\ll r_0$.
This enables us to write
\begin{equation}\label{Eq:g_nn_nn_l}
   \begin{split}
   g^{l}_{n'n'',n_1 n_2}=g_l\left(\delta_{n_1 n_2}\delta_{n' n''}\tilde{g}^{l}_{n'n',n_1 n_1}+
   \delta \tilde{g}^{l}_{n'n'',n_1 n_2}\right)\;,
   \end{split}
\end{equation}
where
\begin{equation}\label{Eq:g_l}
   \begin{split}
   g_l= \frac{1}{2l+1}\frac{1}{r_0},
   \end{split}
\end{equation}
$\tilde{g}^{l}_{n'n',n_1 n_1}=g^{l}_{n'n',n_1 n_1}/g_l\approx 1$
and $\delta \tilde{g}^{l}_{n'n'',n_1 n_2}\ll 1$ so that the second
term in Eq.~\eqref{Eq:g_nn_nn_l} can be neglected to a the leading
order of $d/r_0$.

\section{Chemical potential calculation}\label{App:chem}

With $\varepsilon_{nlm}=\varepsilon_{nlm'}\equiv \varepsilon_{nl}$
and $f^0_{nlm}=f^0_{nlm'}\equiv f^0_{nl}$
Eq.~\eqref{Eq:M_nn_lm_lm} can be written as
\begin{equation}\label{Eq:M_nn_lm_lm_degenerate}
      M^{n'n''}_{lm,\tilde{l}\tilde{m}}=-\sum_{l',l''}\frac{f^0_{n'l'}-f^0_{n''l''}}{\varepsilon_{n'l'}-\varepsilon_{n''l''}}
      \sum_{m',m''} y^{l'm'}_{lm,l''m''}
      y^{l'm'}_{l''m'',\tilde{l}\tilde{m}}\;.
\end{equation}
The last sum on the right hand side of this equation we evaluate
using the following property of the Clebsch-Gordan coefficients
\cite{Varshalovich}
\begin{equation}\label{Eq:sum_Clebsh}
    \sum_{m_1,m_2}C^{jm}_{j_1m_1\: j_2m_2}C^{j'm'}_{j_1m_1\:
    j_2m_2}=\delta_{jj'}\delta_{mm'}.
\end{equation}
Thereby  we make use of the  definition \eqref{Eq:y_lm_lm_lm} and
rewrite
\begin{equation}\label{Eq:y_rewrite}
      y^{l'm'}_{lm,l''m''} y^{l'm'}_{l''m'',\tilde{l}\tilde{m}}=
      (-1)^{m+\tilde{m}}y^{l\:-m}_{l'\:-m',l''m''} y^{\tilde{l}\:-\tilde{m}}_{l'\:-m',l''m''}\;.
\end{equation}
Summing over $m'$ and $m''$ and using Eq.~\eqref{Eq:sum_Clebsh} we
find
\begin{equation}\label{Eq:y_sum}
      \sum_{m',m''} y^{l'm'}_{lm,l''m''} y^{l'm'}_{l''m'',\tilde{l}\tilde{m}}=
       \delta_{l,\tilde{l}}\delta_{m,\tilde{m}}K^{l}_{l'l''}\;,
\end{equation}
where $K^{l}_{l'l''}$ is given by Eq.~\eqref{Eq:K}. With this
finding Eq.~\eqref{Eq:M_nn_lm_lm_degenerate} simplifies to
\begin{equation}\label{Eq:M_nn_lm_lm_simplified}
      M^{n'n''}_{lm,\tilde{l}\tilde{m}}=\delta_{l,\tilde{l}}\delta_{m,\tilde{m}}M_l^{n'n''}\;,
\end{equation}
where $M_l^{n'n''}$ is defined by Eq.~\eqref{Eq:M_nn_l}. Equation
\eqref{Eq:n_nn_lm} can be then simplified to
\begin{equation}\label{Eq:n_nn_lm_simplified}
      \delta n^{n'n''}_{l m}=
     M_l^{n'n''} \delta\mu^{n'n''}_{lm}\;,
\end{equation}
and after dividing this by $M_l^{n'n''}$ we get finally
Eq.~\eqref{Eq:mu_nn_simplified}.

\section{Approximate stationary solutions}\label{App:approx_stat_sol}
Utilizing the spherical shell approximation  ($d/r_0\ll 1$) we
find $\tilde{g}_{nn,n_1n_2}\ll 1$ and $\tilde{r}_{n_1n_2}\ll 1$
for $n_1\neq n_2$ whereas $\tilde{g}_{nn,n_1n_1}\approx 1$ and
$\tilde{r}_{n_1n_1}\approx 1$. As a consequence it follows from
Eq.~\eqref{Eq:z_n1n2_dynamics} that the intraband components
$z_{11}(\omega,t)$ and $z_{22}(\omega,t)$ of the response function
are only weakly influenced by the interband components
$z_{12}(\omega,t)$ and $z_{12}(\omega,t)$. Therefore, we obtain an
approximate solution by solving the equation system
\eqref{Eq:z_n1n2_dynamics} considering only the intraband terms.
This leads  to a system of two coupled equations for
$z_{11}(\omega,t)$ and $z_{22}(\omega,t)$.

The equilibrium population of levels is determined by the
energetic order of the states. The expressions for $W^{nn}(t,t')$
and $I^{nn}(t,t')$ can thus be written approximately as
\begin{eqnarray}
      W^{nn}(t,t')&=&\omega_{{\rm p},n}^2\:e^{\frac{1}{\tau}(t'-t)}
      \frac{\sin\left[\omega_{{\rm F},n}(t'-t)\right]}{\omega_{{\rm F},n}},\label{Eq:W^1_tt}\\
      I^{nn}(t,t')&=&\frac{1}{\tau}e^{\frac{1}{\tau}(t'-t)}\cos\left[\omega_{{\rm F},n}(t'-t)\right],\label{Eq:I^1_tt}
\end{eqnarray}
where
\begin{equation}\label{Eq:w_p_n}
    \omega_{{\rm p},n}=\sqrt{\frac{\omega_{{\rm F},n}}{\omega_{{\rm F},n}-E_{{\rm
    g}}} \frac{e^2N_n}{\epsilon_0 m_0}\frac{1}{6\pi
    r_0^3}}\; .
\end{equation}
 $N_n$ is the number of electrons in the $n$-th radial
subband. $\hbar\omega_{{\rm F},n}$ are the energy distances
between the highest occupied state in the $n$-th radial subband
and the unoccupied state in the same subband having the value of
$l$ that is greater by one. This energy distance can be written as
\begin{equation}\label{Eq:w_F_n}
    \hbar\omega_{{\rm F},n}=\sqrt{2N_n}\frac{\hbar^2}{2m_0r_0^2}+E_{{\rm
    g}}\;,
\end{equation}
where the first term on the right hand side determines the
corresponding energy gap between the highest occupied state and
the lowest unoccupied state for $N_n$ free electrons in a thin
spherical layer with the radius $r_0$ and $E_{{\rm g}}$ is an
additional band gap appearing for the real ``spherical'' molecule
C$_{60}$.

With kernels determined by Eqs.~\eqref{Eq:W^1_tt} and
\eqref{Eq:I^1_tt} the system of integral equations for
$z_{11}(\omega,t)$ and $z_{22}(\omega,t)$ can be reduced to two
coupled ordinary differential equations, each of which describes
the plasmon dynamics in the respective radial band:
\begin{equation}\label{Eq:z_intraband_ODE1}
     \begin{split}
    \frac{{\rm d}^2}{{\rm d}t^2}{z}_{11}(\omega,t)&+\left(\frac{1}{\tau}\!-\!2i\omega\right)\frac{{\rm d}}{{\rm
   d}t}{z}_{11}(\omega,t)\\
     &+\left(\omega_{{\rm p},1}^2\tilde{g}_{11,11}\!+\!\omega_{{\rm
     F},1}^2\!-\!\omega^2\!-\!i\frac{\omega}{\tau}\right)z_{11}(\omega,t)\\
     &+\omega_{{\rm p},1}^2\tilde{g}_{11,22}z_{22}(\omega,t)=-\omega_{{\rm p},1}^2\;,
 \end{split}
\end{equation}
\begin{equation}\label{Eq:z_intraband_ODE2}
     \begin{split}
    \frac{{\rm d}^2}{{\rm d}t^2}{z}_{22}(\omega,t)&+\left(\frac{1}{\tau}\!-\!2i\omega\right)\frac{{\rm d}}{{\rm
   d}t}{z}_{22}(\omega,t)\\
     &+\left(\omega_{{\rm p},2}^2\tilde{g}_{22,22}\!+\!\omega_{{\rm
     F},2}^2\!-\!\omega^2\!-\!i\frac{\omega}{\tau}\right)z_{22}(\omega,t)\\
     &+\omega_{{\rm p},2}^2\tilde{g}_{22,11}z_{11}(\omega,t)=-\omega_{{\rm
p},2}^2\;.
 \end{split}
\end{equation}
Here we made use of the fact that the corrections to
$\tilde{r}_{11}= \tilde{r}_{22}= 1$ are of a second order in
$d/r_0$ and can be neglected.
Then the time- and frequency-dependent dipolar polarizability
is expressed as
\begin{equation}\label{Eq:alpha_via_z1_z_2}
    \alpha(\omega,t)= -r_0^3\left[z^{11}(\omega,t)+z^{22}(\omega,t)\right]\;.
\end{equation}

The stationary solution of the system \eqref{Eq:z_intraband_ODE}
can be found as
\begin{equation}\label{Eq:z_intraband_ODE_stationary1}
     \begin{split}
      z_{11}(\omega)=\frac{1}{D}\bigg[&\omega_{{\rm p},1}^2\omega_{{\rm
         p},2}^2\tilde{g}_{11,22}\\
              &-\omega_{{\rm p},1}^2\left(\omega_{{\rm p},2}^2\tilde{g}_{22,22}+\omega_{{\rm F},2}^2-\omega^2-i\frac{\omega}{\tau}
              \right)\bigg]\;,
      \end{split}
\end{equation}
\begin{equation}\label{Eq:z_intraband_ODE_stationary2}
   \begin{split}
     z_{22}(\omega)=\frac{1}{D}\bigg[&\omega_{{\rm p},1}^2\omega_{{\rm
         p},2}^2\tilde{g}_{11,22}\\
         &-\omega_{{\rm p},2}^2\left(\omega_{{\rm p},1}^2\tilde{g}_{11,11}+\omega_{{\rm F},1}^2-\omega^2-i\frac{\omega}{\tau}
         \right)\bigg]\;,
     \end{split}
\end{equation}
where $\tilde{g}_{11,22}=\tilde{g}_{22,11}$ and
\begin{equation}\label{Eq:D}
 \begin{split}
     D=&
   \left(\omega_{{\rm p},2}^2\tilde{g}_{22,22}+\omega_{{\rm
   F},2}^2-\omega^2-i\frac{\omega}{\tau}\right)\\
     &\times\left(\omega_{{\rm p},1}^2\tilde{g}_{11,11}+\omega_{{\rm
     F},1}^2-\omega^2-i\frac{\omega}{\tau}\right)\\
   &-\omega_{{\rm p},1}^2\omega_{{\rm p},2}^2\left(\tilde{g}_{11,22}\right)^2\;.
 \end{split}
\end{equation}
The stationary frequency-dependent dipolar polarizability  is
given then by
\begin{equation}\label{Eq:z_stationary}
   \alpha(\omega)=-r_0^3\left[z_{11}(\omega)+z_{22}(\omega)\right]\:.
\end{equation}

The plasmon frequencies as retrieved  from
Eq.~\eqref{Eq:z_stationary} derive from the zeros of  denominator
(for  $1/\tau=0$). Since $\omega_{{\rm F},1}^2-\omega_{{\rm
F},2}^2 \ll \omega_{{\rm p},1}^2\tilde{g}_{11,11}-\omega_{{\rm
p},2}^2\tilde{g}_{22,22}$ we find two possible frequencies
$\omega_+$ and $\omega_-$ such that
\begin{eqnarray}
    \hspace{-0.8cm}\omega_+^2&=&\omega_{\rm p}^2+\frac{1}{\omega_{\rm p}^2}
    \left(\omega_{{\rm F},1}^2\omega_{{\rm p},1}^2\tilde{g}_{11,11}
       +\omega_{{\rm F},2}^2\omega_{{\rm p},2}^2\tilde{g}_{22,22}\right)\!,\label{Eq:w_plus}\\
    \hspace{-0.8cm}\omega_-^2&=&\frac{1}{\omega_{\rm p}^2}
    \left(\omega_{{\rm F},1}^2\omega_{{\rm p},2}^2\tilde{g}_{22,22}
       +\omega_{{\rm F},2}^2\omega_{{\rm p},1}^2\tilde{g}_{11,11}\right)\!,\label{Eq:w_minus}
\end{eqnarray}
where
\begin{equation}\label{Eq:w_p}
    \omega_{\rm p}^2=\omega_{{\rm p},1}^2\tilde{g}_{11,11}+\omega_{{\rm
    p},2}^2\tilde{g}_{22,22}\;.
\end{equation}
The amplitude of the peak centered around the  frequency
$\omega_-$ has much smaller value because it contains terms of the
order
 $(\omega_{{\rm F},1}^2-\omega_{{\rm F},2}^2)/(\omega_{{\rm
p},1}^2\tilde{g}_{11,11}-\omega_{{\rm p},2}^2\tilde{g}_{22,22})$.
From Eq.~\eqref{Eq:w_p} we calculate $\omega_+=25.4$~eV and
$\omega_-=8$~eV. Comparing the results found so with those without
making the above approximations confirms the  accuracy of the
approximate expressions for the peak positions.

\section{Time- and frequency-dependent dipolar polarizability}\label{App:dipolar_polarizability}
Given the time-dependent perturbation of the system density
$\delta n(\vec{r},t)$ the induced dipole moment $\vec{P}(t)$ is
determined then by
\begin{equation}\label{Eq:induced_dipole}
    \vec{P}(t)=e\int  \vec{r}\: \delta n(\vec{r},t)\: {\rm d}^3
    \vec{r}\;.
\end{equation}
Inserting here Eqs.~\eqref{Eq:delta_n_full} and
\eqref{Eq:delta_n_lm} we find
\begin{equation}\label{Eq:induced_dipole_lm_nn}
    \vec{P}(t)=e\int  \vec{r}\:  \sum_{n'n''} s_{n'n''}(r) \sum_{lm} \delta n^{n'n''}_{lm}(t) Y_{lm}(\theta,\phi)\: {\rm d}^3
    \vec{r}\;.
\end{equation}
The components $\delta n^{n'n''}_{lm}(t)$ can be calculated once
we have the components $V_{lm}^{{\rm ext},n' n''}(t)$ of the
external perturbation and the response function $\chi^l_{n'
n'',n_1n_2}(t,t')$ using Eq.~\eqref{Eq:chi_general_def}.

If we consider the response of the system to an external spatially
homogenous time-dependent electric field $\vec{{\cal E}}(t)$, it
is enough to consider only the dipolar components of the response
function, i.e. the components with $l=0$. We may select the
$z$-axis to be parallel to the electric field. Then the external
potential reads:
\begin{equation}\label{Eq:V_ext_z}
   V^{\rm ext}(\vec{r},t)=-ez {\cal E}(t)\;.
\end{equation}
Considering the angular components of this potential we conclude
that only the component with $l=1$ and $m=0$ does not vanish:
\begin{equation}\label{Eq:V_ext_z_lm}
   V^{\rm ext}_{10}(\vec{r},t)=-2\sqrt{\frac{\pi}{3}} er {\cal E}(t)\;.
\end{equation}
As a consequence in Eq.~\eqref{Eq:induced_dipole_lm_nn} only terms
with $l=1$ and $m=0$ survive and we infer that the induced dipole
$\vec{P}(t)$ moment is directed parallel to the electric field
$\vec{{\cal E}}(t)$. Its magnitude is given by
\begin{equation}\label{Eq:induced_dipole_t}
  \begin{split}
    P(t)=-\frac{4\pi}{3}e^2r_0^2   \int_{-\infty}^{t}\!\! {\rm
    d}t'\!\!
    \sum_{n'n'',n_1n_2} &\tilde{r}_{n'n''} \chi_{n' n'',n_1n_2}(t,t')\\
    &\times\tilde{r}_{n_1n_2} {\cal E}(t')\;,
  \end{split}
\end{equation}
where
\begin{equation}\label{Eq:r_nn}
    \tilde{r}_{n'n''}=\frac{1}{r_0}\int\!\! r^2 {\rm d}r\: s_{n'n''}(r) r\;.
\end{equation}
For a perturbation with an electric field ${\cal E}(t)={\cal
E}_0{\rm e}^{-i\omega t}$ at a particular frequency $\omega$ we
derive  using Eq.~\eqref{Eq:z_n1n2_def} the expression
\begin{equation}\label{Eq:induced_dipole_omega}
    P(\omega;t)=-4\pi r_0^3\epsilon_0\sum_{n'n''} \tilde{r}_{n'n''} z_{n'n''}(\omega,t) {\cal
    E}_0{\rm e}^{-i\omega t}\;.
\end{equation}
The time- and frequency-dependent dipolar polarizability
$\alpha_{\rm SI}(\omega,t)$ in SI units is then found to be equal
to
\begin{equation}\label{Eq:alpha_omega_t_SI}
    \alpha_{\rm SI}(\omega,t)=-4\pi \epsilon_0 r_0^3\sum_{n'n''} \tilde{r}_{n'n''} z_{n'n''}(\omega,t)\;.
\end{equation}
In literature, however, the polarizability is more frequently
expressed as polarizability volume (as it happens naturally when
using Gauss units): $\alpha=\frac{1}{4\pi\epsilon_0}\alpha_{\rm
SI}$. Using this definition we end up with
\begin{equation}\label{Eq:alpha_omega_t}
    \alpha(\omega,t)=- r_0^3\sum_{n'n''} \tilde{r}_{n'n''} z_{n'n''}(\omega,t)\;.
\end{equation}
We  note that  the factors $z_{n'n''}(\omega,t)$ are
dimensionless.

For  $d/r_0\ll 1$ and considering only the lowest order
contribution we conclude $\tilde{r}_{n'n''}\approx
\delta_{n',n''}$ and therefore
\begin{equation}\label{Eq:alpha_omega_t_surface}
    \alpha(\omega,t)\approx - r_0^3 \sum_{n}
    z_{nn}(\omega,t)\;.
\end{equation}
With this expression we can describe the approximate response
given only by the intraband excitations.


%

\end{document}